# Threshold Logic Computing: Memristive-CMOS Circuits for Fast Fourier Transform and Vedic Multiplication

Alex Pappachen James, Dinesh S. Kumar, and Arun Ajayan

***Abstract*—Brain inspired circuits can provide an alternative solution to implement computing architectures taking advantage of fault tolerance and generalisation ability of logic gates. In this brief, we advance over the memristive threshold circuit configuration consisting of memristive averaging circuit in combination with operational amplifier and/or CMOS inverters in application to realizing complex computing circuits. The developed memristive threshold logic gates are used for designing FFT and multiplication circuits useful for modern microprocessors. Overall, the proposed threshold logic outperforms previous memristive-CMOS logic cells on every aspect, however, indicate a lower chip area, lower THD, and controllable leakage power, but a higher power dissipation with respect to CMOS logic.**

## I. Introduction

There has been several attempts to functionally and electrically mimic the neuron activity and its networks [1]. However, major challenge that deters the progress in VLSI implementations of brain like logic gates is the scalability of the networks and its practical limitations in solving large variable boolean logic problems. The possibility to mimic the brain like circuits and logic networks is a topic of intense debate. One possibility is to apply the threshold logic gates in designing conventional computational blocks, while other option is to develop a completely trainable architecture that does not strictly bind itself to conventional computing topologies. In this brief, we restrict the notion of the brain mimicking to develop a generalised memristive threshold logic cell in application to designing conventional computing blocks. Nonetheless, this topic is one of the forefront challenges in development of on-chip brain computing, and would require us to investigate not just new circuit design logic, but also new devices and systems.

Threshold logic is the primary logic of human brain that inspires from the neuronal firing and training mechanisms. The progress in threshold logic circuits [2] are often limited to implementation of logic gates with few number of input variables, this leads to limited progress being made in the development of practical computing circuit topologies.

Memristor like switching devices [3] unlike many other electronic devices has an interesting appeal in on-chip brain computing, as it offers switching state through its bi-level resistance values. Further these resistors are mapped to the binary memory space and offer the advantage of low on-chip area and low leakage currents. We explore this aspect of memristor, and extend over our previous work [4] in designing FFT computing useful for signal processing applications, and vedic additions and multiplications for efficient ALU design. The resulting circuits can be used in combination with conventional CMOS circuits to develop threshold logic processor designs.

## II. Computing Circuits with Threshold Logic

The memristive threshold logic (MTL) cell shown in Fig. 1 is the basic cell which consists of two parts; a memristor based input voltage averaging circuit and an output threshold circuit. In contrast to our previous work on resistive threshold logic [4], the proposed cell has the input potential divider circuit modified by removing the pull-down resistor to form an input voltage averaging circuit and the threshold circuit modified with the combined use of operational amplifier and CMOS inverter. In particular, by removing the pull-down resistor from [4], an important improvement over lower power dissipation is achieved in cell as shown in the Table I.

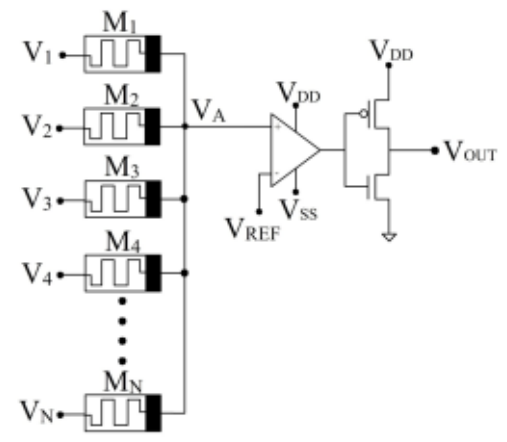

Fig. 1: Memristive threshold logic cell

In the presented work, the threshold unit consisting of a combination of an op-amp [5] and a CMOS inverter that allows for fault tolerance in terms of logical output signal stability. The generalization of the cell to work as different logic gates is achieved with the ability of the cell to utilize a wider range of threshold value.

TABLE I: Comparing Previous Circuit [4] with present Proposed Circuit by Implementing a 2-input NOR Logic Gate.

| Logic Family | Power dissipation (μW) |
|---|---|
| RTL without op-amp [4] | 8.30 |
| MTL without op-amp | 3.00 |
| RTL with op-amp [4] | 19.70 |
| MTL with op-amp | 16.61 |

For an $N$ input cell, the resistance circuit part consist of $N$ memristors having equal memristance values, $M_1 = M_2 = ...M_N = M$. The output voltage $V_A$ for $N$ input voltages $V_I$ can be represented as $V_A = (\sum_{I=1}^{N} V_I)/N$. Table II shows truth table for two input NAND and NOR gates. $V_1$ and $V_2$ are the input voltage that can take values of $V_L$ (voltage low) or $V_H$ (voltage high). For practical purposes, the boundary conditions are avoided, and in general for any N inputs, if $V_{REF}$ is in between $((N-1)V_H + V_L)/N$ and $V_H$ we obtain NAND logic and if it is in between $V_L$ and $(V_L + (N-1)V_H)/N$ we obtain NOR logic. The combined effect of $V_{REF}$ at operational amplifier and $V_{TH}$ of the inverter provides a stable threshold logic unit, where $V_{TH}$ is the threshold voltage of the inverter.

The operational amplifier ensures a wider range of threshold value limiting the role of inverter as a means to ensure stable binary states. The impact of having operational amplifier in the output of the cell is shown in Fig. 2a, while Fig. 2b captures the variations of the output voltage for different values of $V_{REF}$. The advantages of using op-amp in order to fix the threshold of the circuit can be clearly observed. from the Fig. 2. Only for the gates that having inputs higher than 2 input require the use an op-amp in the circuit. In addition to this It has been observed during simulation that, without using an op-amp in the circuit and a +/-15% variability in channel lengths does not have any effect on the output of NAND logic, while minor variations in NOR logic. On the contrary, we observed no variation in outputs even with +/-15% variability in channel lengths when op-amps are incorporated. The reason for this is because the amplification in the voltage range made by the op-amp increases the voltage range to +/-1. This offers broader selection range of CMOS inverters thresholds ensuring that the threshold values lie inside this range even if the

A.P. James is a faculty with Electrical and Electronic Engineering department, Nazarbayev University. D.S. Kumar and A. Ajayan are member research staff with Enview R&D labs. Contact Email: apj@ieee.org

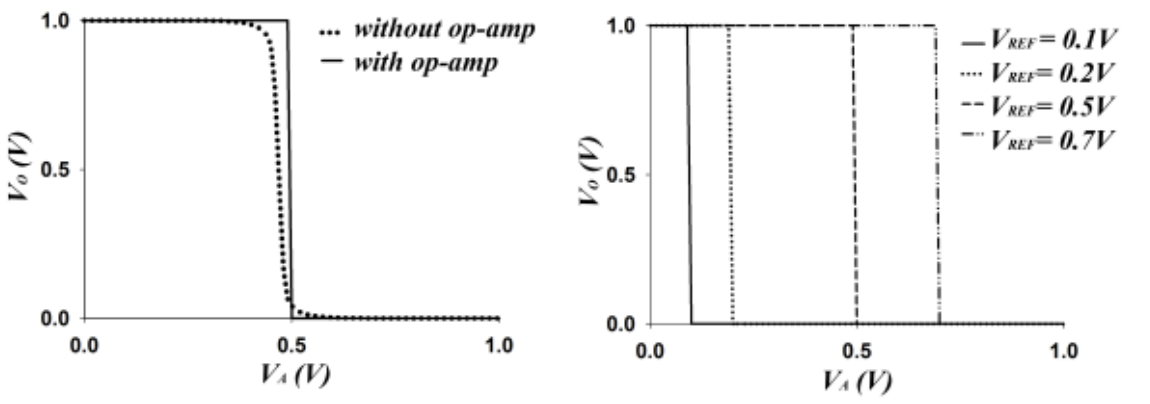

(a) (b)

Fig. 2: Relation between output of the memristive divider and output of the inverter of the proposed cell. (a) shows the effect of op-amp on the output of the cell, and (b) shows the variation in output of the cell for different values of $V_{REF}$.

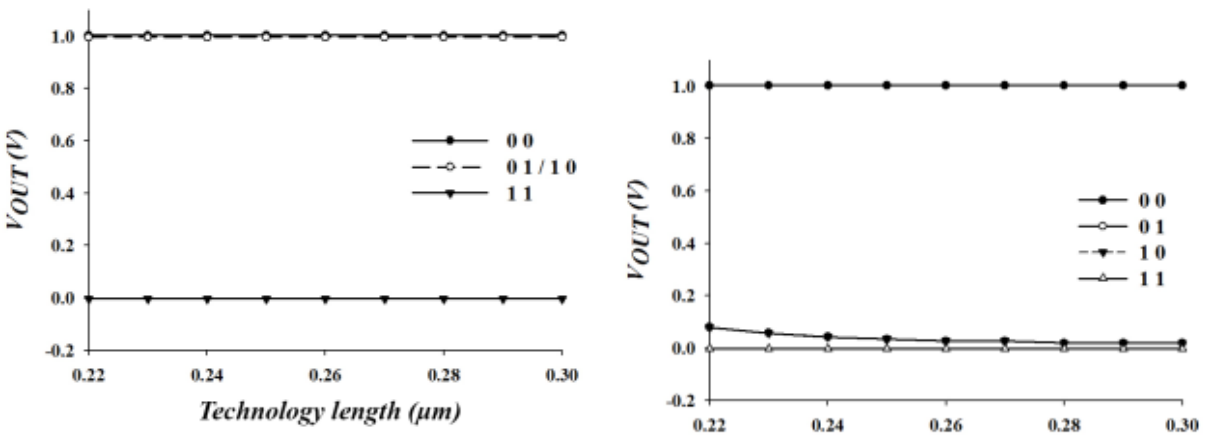

(a) NAND (b) NOR

Fig. 3: The graphs illustrate the effect of change in the technology length on the output of MTL logic family without op-amp.

technology length are changed. Additionally change in the speed of the transistors will reflect in the delay introduced by inverter, i.e., slow-slow (0.89ns), fast-fast (0.23ns), slow-fast (0.50ns), fast-slow (0.52ns) tested with an input pulse with speed of 10ns. The delay is high for the slow-slow condition and less in-case of fast-fast.

Fig. 3 shows the effect of change in technology length on the output of MTL logic gates. While checking the effect of other process variations, temperature and chain of logic gates, it is observed that the power dissipation has a linear change in accordance with the temperature change and has no effect by the chain of logic gates. In case of delay, change in temperature doesn't have any significant effect but the chain of logic gates will increase the delay linearly with an increase of $d_1$ in each level, where $d_1$ is the delay of a single cell. For this study, inverter configuration of the proposed cell is used and a chain of 6 inverters are checked in order to get the effect of chain of logic gates. The results are shown in the Fig. 4.

Throughout this paper, we use the non-ideal resistive switching model of memristor reported in [4] for our study with an area of 10nm $\times$ 10nm and resistances in the range of $[10^{-6}\Omega, 10^{-12}\Omega]$, while CMOS circuits uses 0.25 µm (both in the MTL and CMOS logics)

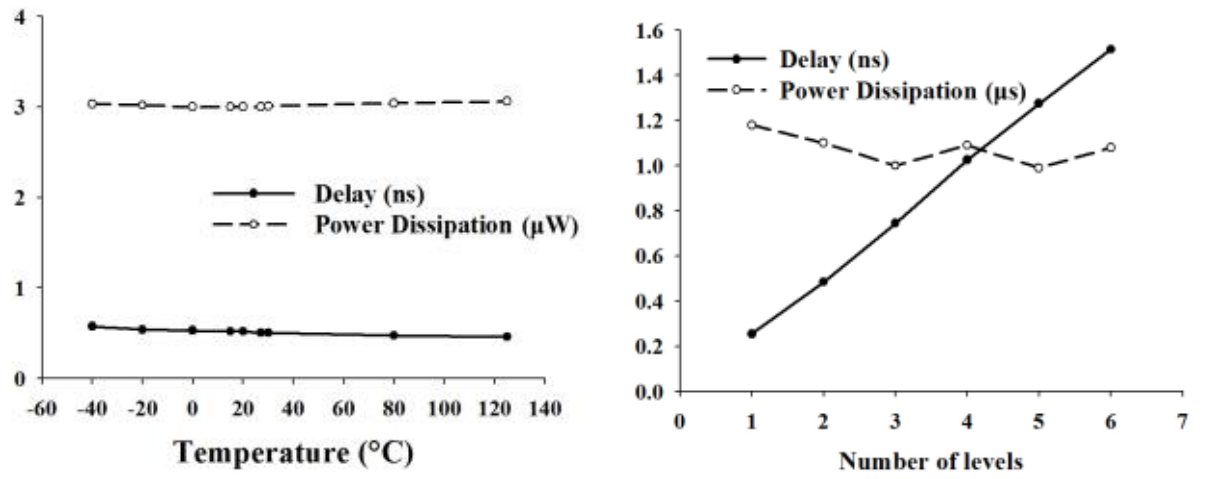

(a) (b)

Fig. 4: The graphs show the effects of (a) changes in temperature, and (b) delay introduced by chain of logic gates on power dissipation and delay.

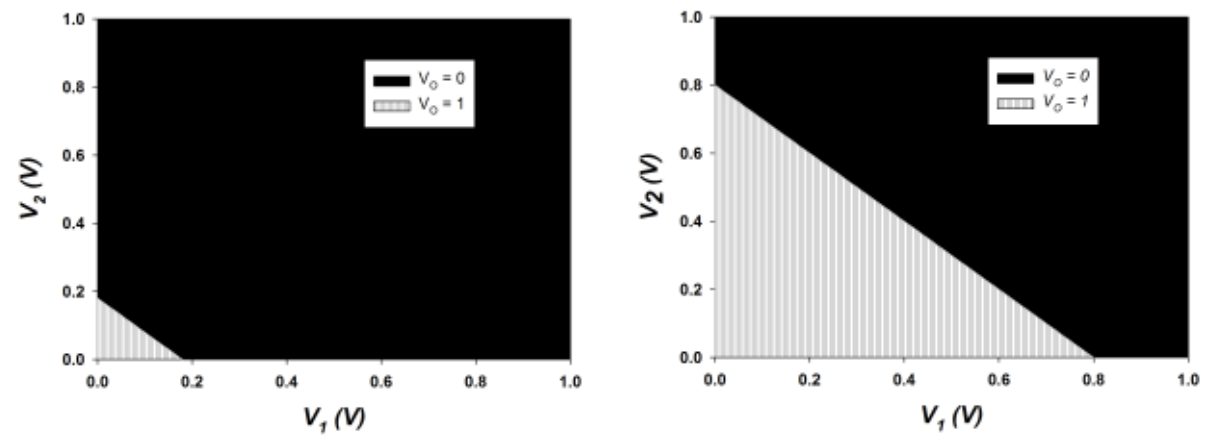

(a) (b)

Fig. 5: Operating range of two input (a) NOR cell (b) NAND memristive threshold logic cell with $V_{REF} = 0.5$V .

TSMC (Taiwan Semiconductor Manufacturing Company) technology to reflect the practical applicability in standard silicon technologies. The SPICE (Simulation Program with Integrated Circuit Emphasis) models take into account the extracted parasitics from CMOS layout, so as to ensure the accuracy for practical realisation. Memristor model has a non-ideal behaviour and take into account the boundary effects. In practise, the memristor and CMOS layers can be separated into different layers [6], [7]. Memristors can be fabricated directly above the CMOS circuits by making them as a part of the interconnect. A 2D array of vias provides electrical connectivity between the CMOS and the memristor layers. Since the variability between the memristors are practically limited and there is a large threshold range for $V_{REF}$ for achieving desired logical functionality, the errors resulting from memristance tolerance can be ignored.

TABLE II: TRUTH TABLE FOR THE MEMRISTIVE THRESHOLD LOGIC CELL

| $V_1$ | $V_2$ | $V_A$ | NOR [1] | NAND [2] |
|---|---|---|---|---|
| $V_L$ | $V_L$ | $\frac{V_L+V_L}{2}$ | $V_H$ | $V_H$ |
| $V_L$ | $V_H$ | $\frac{V_L+V_H}{2}$ | $V_L$ | $V_H$ |
| $V_H$ | $V_L$ | $\frac{V_H+V_L}{2}$ | $V_L$ | $V_H$ |
| $V_H$ | $V_H$ | $\frac{V_H+V_H}{2}$ | $V_L$ | $V_L$ |

[1] N input NOR threshold: $V_L < V_{REF} < \frac{(N-1)V_L+V_H}{N}$
[2] N input NAND threshold: $\frac{(N-1)V_H+V_L}{N} < V_{REF} < V_H$

Fig. 5 shows the operating range for NOR and NAND logic for two input cell where $V_H = 1$V and $V_L = 0$V. It is observed that the gate provides robust functional performance even when there is an input signal variability of 20%. Figure 6a shows the input and output signal waveform of a two input NOR cell. Figure 6b shows the circuit diagrams of OR, AND and XOR functions implemented using the proposed NOR logic cell, where the $V_{REF} = V_L + \delta$ with $\delta$ representing the incremental threshold value required for the functional implementation of threshold logic cell. As the number of input increases the voltage range in which a threshold can be fixed will get narrow. Hence for each number of inputs the threshold value $V_{REF}$ have to be fixed separately. In order to avoid this problem a $V_{REF}$ value close to $V_L$ and $V_H$ for NOR and NAND configuration should be selected. $V_L + \delta$ is a voltage value that is close to $V_L$ and less than $((N-1)V_L + V_H)/N$. This will give the freedom of using the cell without changing $V_{REF}$ for increased number of inputs.

Total harmonic distortion (THD) of the NOR and NAND cells are calculated using test inputs whose signal frequency changes from 20Hz to 20MHz. The comparison of the THD results with other logic families is shown in Fig. 7a. It is observed that the memristive threshold logic has better immunity to THDs when compared to other logic families. This implies an improved accuracy and simplicity of circuit design for timing sensitive digital circuit applications.

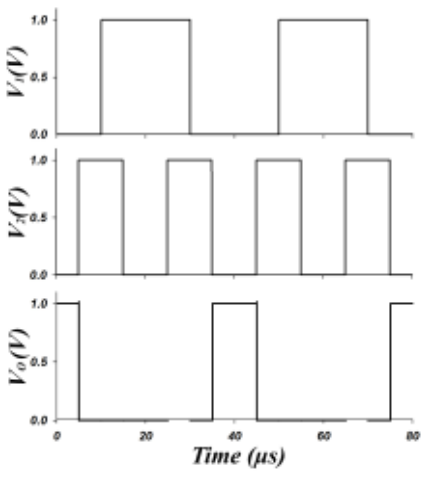

(a)

(b)

Fig. 6: (a) Input and output signal waveforms of the two input proposed cell with op-amp in NOR logic (b) Circuit diagrams of the logic gates using proposed cell - (i) OR gate (ii) AND gate (iii) XOR gate.

(a)

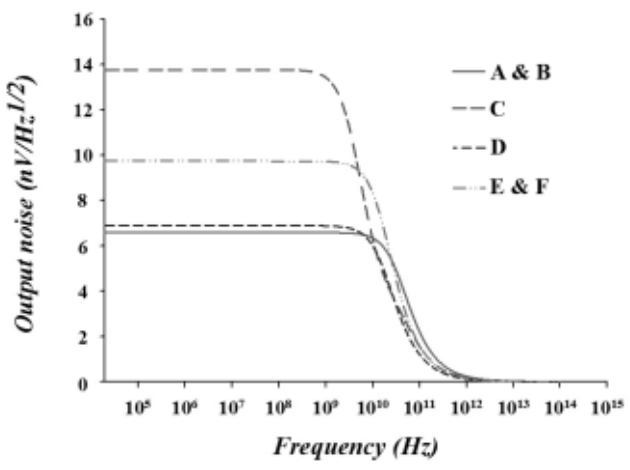

(b)

Fig. 7: (a) Total harmonic distortion (b) Noise spectral density, of A - MTL NAND, B - MTL NOR, C - CMOS NAND, D - CMOS NOR, E - PSEUDO NAND, F - PSEUDO NOR, G - DYNAMIC CMOS NAND, and H - DYNAMIC CMOS NOR

Fig. 7b shows that the noise spectral density of the MTL logic is better than the other two technologies. From THD and noise comparison we can conclude that the proposed MTL logic family has more immunity to noise and harmonic distortion.

Table III shows the example comparison of the memristive threshold logic for NOR logic gate with CMOS logic family for area, power dissipation, energy and leakage power. It can be seen that even for low number of inputs the area and leakage power is lower than CMOS, while the power dissipation and energy is higher due to the use of CMOS and operational amplifier circuits for threshold design. It may be noted RTL and MTL enable the possibility of large number of inputs as demonstrated in [4] and can reduce the area requirements and leakage power significantly for implementing digital logic circuits.

TABLE III: Comparison of Area, Power Dissipation, Leakage Power and Energy of the 2-input Cell with other Logic Families in NOR Configuration.

| Logic Family | Area (µm$^2$) | Power dissipation(W)[a] | Leakage Power (W) | Energy[a] (J) |
|---|---|---|---|---|
| CMOS | 9.4 | 28.6p | 16.32p | 28.6z |
| MTL (without op-amp) | 4.55 | 3.00µ | 14.30p | 0.30p |
| MTL (with op-amp) | 31.30 | 19.70µ | 80.96p | 1.09p |

[a] The values for MTL with op-amp are obtained at a maximum speed of 10MHz and for MTL (without op-amp) & CMOS logics the values are obtained at a max speed of 1GHz.

The area given in Table III and mentioned in other tables are based on the device length and width without consideration to wires and interconnects. Since, the physcial layout area can vary significantly based on the optimal layout configurations, it will be difficult to compare the technologies based on the layout in a quantified manner. For example, the on-chip area for a 2 input proposed NOR cell in ElectricVLSI resulted in an area of 95 µm$^2$ for MTL gate, while for CMOS gate this was 125 µm$^2$, when using 0.25µm technology. Because the physical design of the cells would require additional optimisation for the area, through out this paper, we use only aspect ratios of the devices to make the comparison between the circuits from different logic families.

*Other memristive-CMOS threshold logic gates:* There do not exist FFT or multiplier based circuit implementations with memristive threshold logic circuits that can be fairly compared with the example circuits reported in this paper. However, there do exists memristor-CMOS based threshold logic cells that are close counterparts to the proposed cell, that could be used to implement similar computing circuits. For example, RTLG [8] and EEMTL [9] uses reconfigurable architectures that can be used to develop circuits for digital logic problems by changing the memristance of the memristor, and requires additional programming circuits leading to larger area requirements and higher power dissipation than the proposed cell. RTLG uses two op-amps in a single cell, one for programmable part and another for thresholding part. In addition, the programming require pulse generation units based on FPGA and DSP processor, and related interfacing circuits. On the other hand, for EEMTL, in-order to implement a N input cell requires N memristors and (2×N+8) transistors, while in the proposed cell this is N memristors and 12 transistors (maximum). In the proposed cell, the number of transistors will not increase as per the number of inputs, and is one of the advantage over other threshold logic cells such as EEMTL. The comparison between proposed method, RTLG and EEMTL based on the number of transistors required to implement different logic gates is given in the Table IV. However, since these cell technologies are not mature yet for large scale implementations, a fair comparison for larger circuits build with threshold logic is done only with established technologies such as CMOS.

TABLE IV: Comparison of Transistor Count Between RTLG, EEMTL and MTL Logics

| Boolean Function | RTLG [a,b] | EEMTL | MTL[a] |
|---|---|---|---|
| NOR / NAND (N input) | 24 | 2×N+8 | 10 |
| OR / AND (N input) | 24 | 2×N+8 | 12 |

[a] For the fairness in comparison op-amp used in all logics have same number of transistors.

[b] Number of transistors projected is for implementing a bi-level hard threshold function. This number can significantly increase based on the complexity of the threshold function used.

The use of the proposed memristive logic circuits in designing conventional logic circuit is demonstrated using two examples. The first example reports the design of a FFT circuit using memristive threshold logic, while the second example reports the design of a multiplier circuit. Both these circuit configurations are useful in processors for computing purposes.

*Memristive Threshold FFT circuit:* FFT/IFFT is widely used in digital signal processing for various filter implementations. The basic equation of 4-point DFT is $X(k) = \sum_{i=0}^{3} x(n) e^{\frac{-j2\pi nk}{4}}, k = 0, 1, 2, 3$. Using this equation, we can represent the signal flow graph [10] of 4-point DFT as shown in Fig. 8a. Implementation of the FFT processor [11] can be done as shown in Fig. 8b where all the inputs to the circuit are 8 bits long. Having a closer look at the exponential term of a 4-point DFT i.e. $e^{\frac{\mp j2\pi nk}{4}} = \pm 1$ or $\pm \mathbf{j}$, the multiplications with $\pm 1$ and $\pm \mathbf{j}$ are trivial, and no multipliers are needed to implement them. Each FFT unit have 4 inputs and one corresponding FFT output. Inputs are given to the FFT units as shown in Fig. 8a. The inputs which are to be subtracted are complemented and added. From Fig. 8a, it can be seen that the real part and the imaginary part of the first output in 4-point FFT requires only addition operation. Hence the inputs to the first 2 FFT units in Fig. 8b are not complimented. These FFT units are implemented using 3 Carry-Lookahead Adders (CLA) as shown in Fig. 8c.

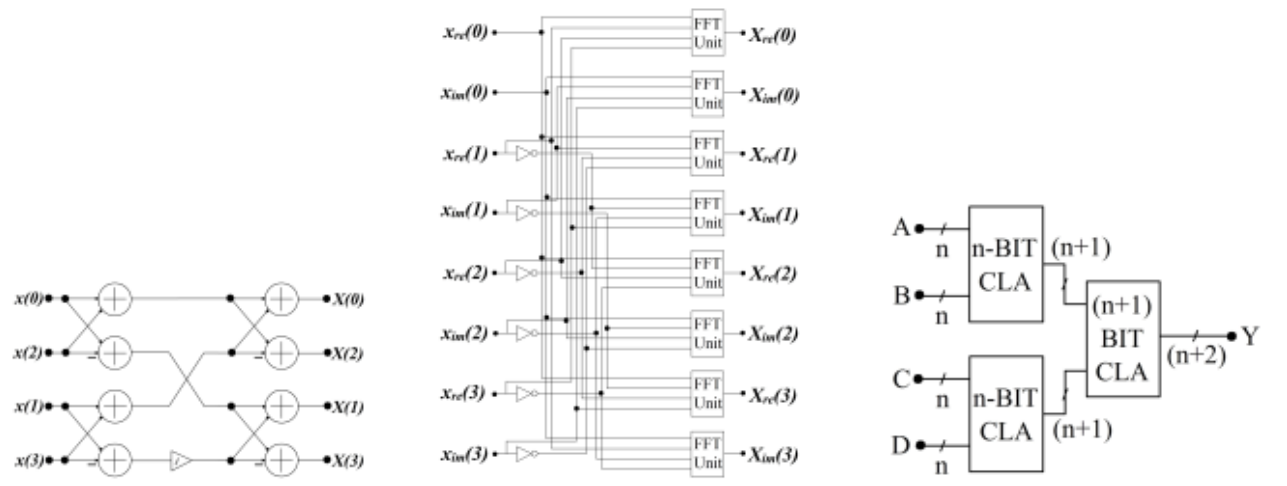

(a) (b) (c)

Fig. 8: (a) Signal flow graph of a 4-point DFT processor (b) Block diagram of a 4 point DFT processor (c) Block diagram of FFT units used in the DFT processor

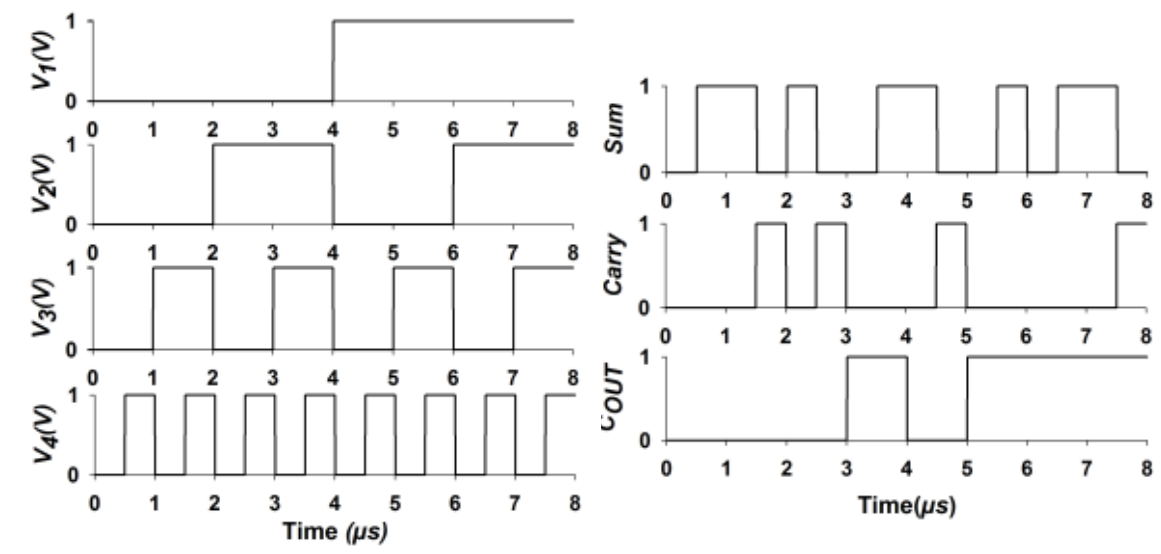

(a) (b)

Fig. 9: (a) $8^{th}$ bit of the 4 inputs to a FFT unit and (b) shows the output sum, carry and $C_{out}$ of the $8^{th}$ bit.

Other than the first 2 FFT units, rest of the six FFT units have 2 additions and 2 subtractions. Inputs to be added are given to the first CLA whereas the inputs that are to be subtracted are complimented and then added using the second CLA. In order to obtain the 2's compliment, 1 is to be added to the LSBs of inverted inputs. For this we utilize the $C_0$ pin of the carry look ahead adders and a logic high is applied to the $C_0$ pin of both the 8-bit CLA. This operation equates to adding one twice. Now, the outputs of these CLAs are added using the third CLA whose output result in the required transform. The Carry-Lookahead adders are implemented using the proposed memristive threshold logic circuits.

Signal waveform of the $8^{th}$ bit from the 4 inputs to the FFT circuit are shown in Fig. 9a, while Fig. 9b shows the least significant bit of the resulting output signals. Table V shows the quantitative performance comparison of the memristive logic and CMOS logic FFT implementations. It is observed that the proposed logic shows reduced area requirement, however, has higher power dissipation. The proposed logic has zero leakage power in its memristor components and the leakage power is entirely contributed by the op-amps. Clearly, the leakage power can be reduced by the designing a low power and low leakage op-amp circuit.

In-order to show the effect on the higher point FFTs, we have implemented a 8-point FFT using vedic multipliers and CLAs with MTL and did the performance comparison with CMOS technology. The results are shown in the Table VI. From the table it can be seen that the advantage of the MTL is its lower on-chip area. The size of the circuit would sufficiently relate to the scalability and performance issue of FFT circuit implementation.

TABLE V: COMPARISON OF AREA, POWER DISSIPATION AND LEAKAGE POWER OF THE PROPOSED FFT CIRCUIT WITH CMOS LOGIC

| Logic Family | Area($\mu m^2$) | Power dissipation(W) | Leakage Power(W) |
|---|---|---|---|
| CMOS | 75942.8 | 11.941n | 6.218n |
| MTL [a] | 42632.4 | 16.156m | 13.46n |

[a] Higher input gates implemented using circuits with Op-amp

TABLE VI: PERFORMANCE COMPARISON OF 8-POINT FFT IMPLEMENTED USING MTL AND CMOS TECHNOLOGIES

| Logic Family | Area($mm^2$) | Power dissipation(W) |
|---|---|---|
| CMOS | 0.4155 | 0.356μ |
| MTL [a] | 0.224 | 79.12m |

[a] Higher input gates implemented using circuits with Op-amp

*Memristive Threshold Vedic Multiplier:* Vedic multiplier is a multiplier architecture that uses vedic mathematic [12] method for its multiplication algorithm. Among the sixteen methods presented in the Vedic Mathematics, due to the parallelism in the mode of operation, we are using Urdhva Thirayakbhyam (vertical and cross-wise method) [13] method for our multiplier architecture. In this technique, all the partial products can be found in parallel and the entire multiplication can be completed by using additional two or three levels of adders.

Based on this algorithm, the architectural block diagram and the working principle for a 2bit multiplier is shown in Fig. 10a. Figure 11a shows the equivalent circuit of a 2 bit vedic multiplier algorithm implemented using memristive threshold logic and Fig. 11b shows the waveform of the corresponding circuit, where $A_0$ $A_1$ and $B_0$ $B_1$ are the 2 bit inputs and $S_0$ $S_1$ $S_2$ and $S_3$ are the four bits of the result. In this algorithm, the 2 bit multiplier is the basic multiplier unit that can be used for making the higher bit multipliers.

In order to implement an $N$ bit memristive threshold vedic multiplier as shown in Fig. 10b, we need four $N/2$ bit multipliers, two $N$ bit CLAs, one $N/2$ bit CLA and a half adder, where $N$ must be in the power of 2. As shown in Fig. 10b, by using four 2 bit multipliers, we can implement 4 bit multiplier. Similarly, by using four 4 bit multiplier, we can implement 8 bit multiplier and will continue the same procedure for any higher bits.

Suppose, the task is to implement a 8 bit multiplication then it would need four 4 bit multiplier, two 8 bit CLAs, one 4 bit CLA and one half adder. First, the 8 bits of both multiplicand and multiplier is divided into two 4 bit numbers. Let A and B be two eight bit numbers, where we divide A as ($A_H$ , $A_L$) and B as ($B_H$, $B_L$). Similar to the steps that we followed in implementing 2 bit multiplier as explained in Fig. 10a, we multiply this four 4 bit numbers ($A_H$, $A_L$, $B_H$ and $B_L$). The partial products are $A_L$ x $B_L$, $A_H$ x $B_L$, $A_L$ x$B_H$ and $A_H$ x $B_H$. Since these are independent operations they are processed in parallel. While doing the multiplication, each of this 4 bit numbers ($A_H$, $A_L$, $B_H$ and $B_L$) will again divide into two 2 bit numbers, that is $A_L$ as $A_{LH}$ and $A_{LL}$, and proceed the 4 bit multiplication as explained in case of 8 bit ($A_L$ x $B_L \rightarrow A_{LL}$ x $B_{LL}$, $A_{LL}$ x $B_{LH}$, $A_{LH}$ x $B_{LL}$, $A_{LH}$ x $B_{LH}$). The basic multiplication unit is a 2 bit multiplier as shown in Fig. 11a.

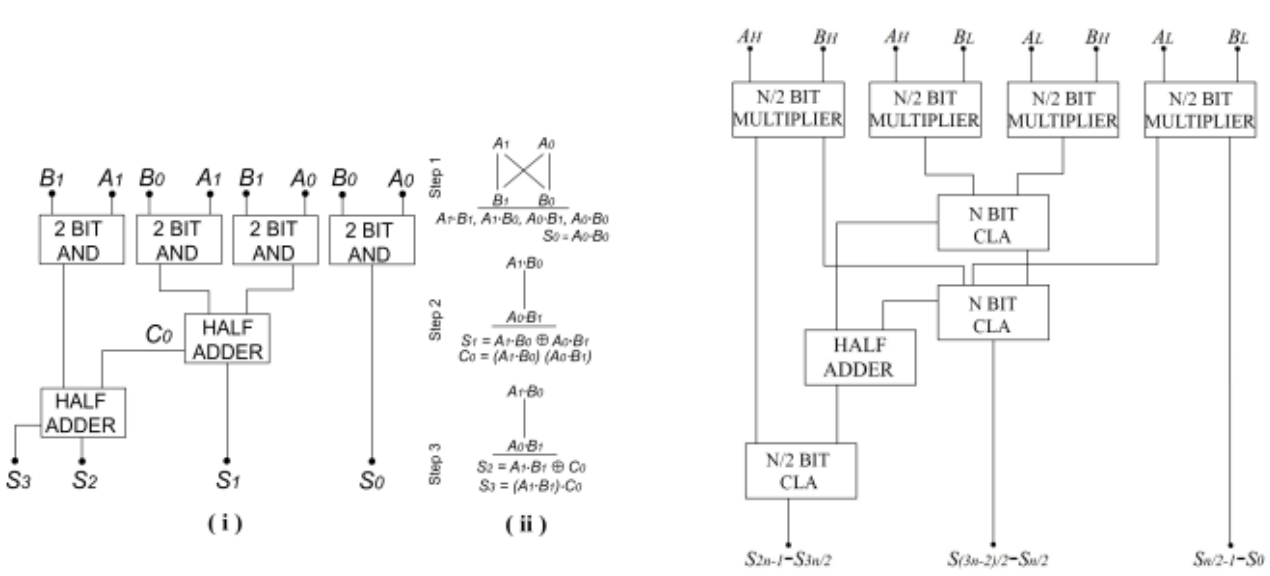

(a) (b)

Fig. 10: (a) Architecture block diagram for a 2 bit vedic multiplier is shown in (i), while (ii) explains the working principle for each stage of the architecture (b) Block diagram for an N bit multiplier

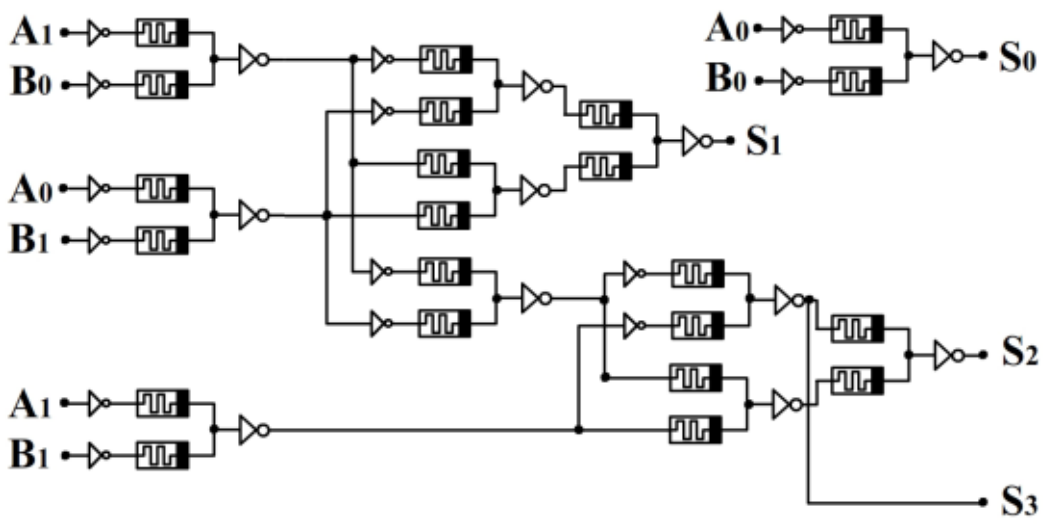

(a)

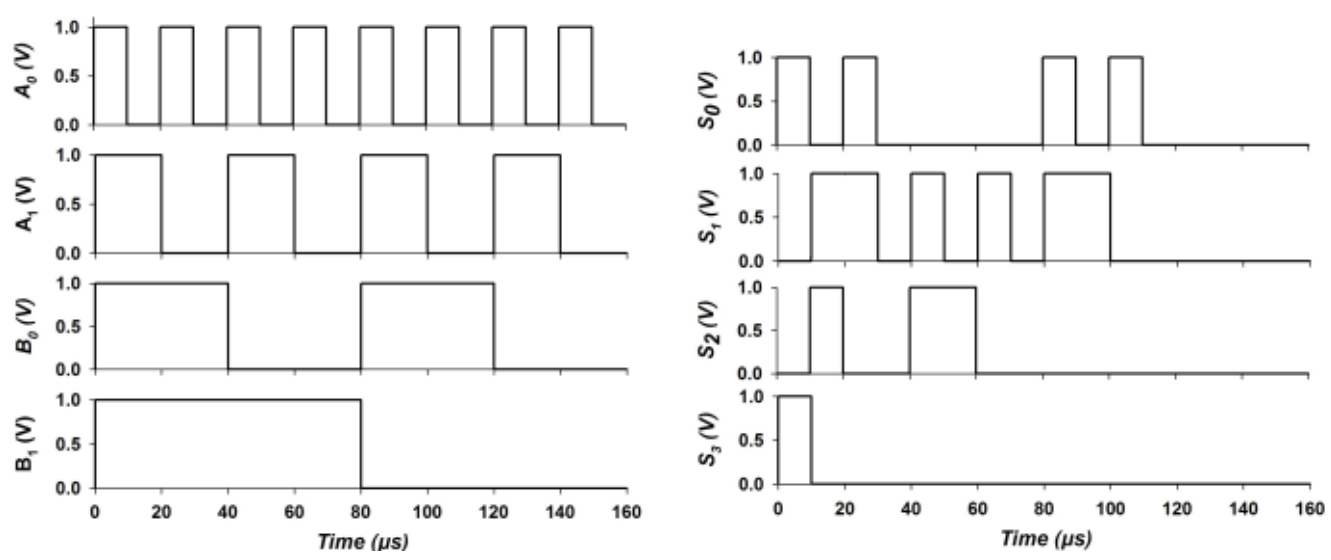

(b)

Fig. 11: 2 bit memristive threshold vedic multiplier (a) circuit (b) input and output waveform of the circuit, where A0 A1 and B0 B1 are the 2 bit inputs and S0 S1 S2 and S3 are the four bits of the output

Table VII shows the comparison of 2 bit and 8 bit vedic multiplier using memristive threshold logic and CMOS logic for area, power dissipation and leakage power. From the table it is clear that the proposed MTL architecture has a clear advantage over the existing CMOS technology. As an initial step to achieve small area on-chip brain computing, this realization is a promising result for the future developments in the field of cognitive computing circuits. For the 2 bit multiplier, all the cells are of two inputs and we implement the circuit with cells without op-amp as shown in Fig. 10a. The op-amp is not required in this case as large variation in threshold is not required to implement the threshold logic. For multiplier with higher number of bits, cells with op-amp are used to ensure tolerance to larger range of threshold values. This will increase the power dissipation as indicated in Table VII. Like the FFT circuit, we expect to overcome this drawback by designing low power op-amps in the circuit.

## III. Conclusion

In this brief, we reported an improved memristance-CMOS threshold logic cell having lower power dissipation and smaller on-chip area footprint. In comparison with CMOS logic the proposed MTL cell implementation have lower area requirements and higher power dissipation, and in comparison with other memristive-CMOS threshold logic gates the proposed cell indicate lower area requirements and lower power dissipation. Further, this brief, reports the successful application of the MTL cells in the examples of FFT and vedic multiplication computing circuits. The MTL cell show robustness to process variability in temperature, memristances and technology lengths indicating the fault tolerance ability of brain like logic circuits. The generalisation ability of the cell, i.e. a single cell structure with multiple functionality is again a characteristic of the proposed logic. The power dissipation and leakage power of the proposed logic is contributed by the op-amp part of the circuit, and can be improved in future by the developing low power high-speed op-amps. However, the design and optimisation of the op-amp that can handle a wide range of threshold with high speed and low power dissipation is left as an open problem for future work.

TABLE VII: Comparison of Overall Area, Power Dissipation and Leakage Power of Proposed Multiplier Circuit with CMOS Logic

| Multiplier | Logic Family | Area($\mu m^2$) | Power dissipation(W) | Leakage Power(W) |
|---|---|---|---|---|
| 2 bit multiplier | CMOS | 169.67 | 0.247n | 0.115n |
| | MTL[a] | 50.05 | 1.09μ | 0.173n[b] |
| 8 bit multiplier | CMOS | 21093.54 | 26.97n | 15.13n |
| | MTL[a] | 9628.55 | 3.01m | 32.19n[c] |

[a] 2 bit multipliers implemented without using op-amp, all higher input gates implemented with op-amp.
[b] Memristor effectively have no leakage power, the given value is that contributed by the CMOS inverter
[c] This value is entirely contributed from the threshold unit (op-amp and CMOS inverter)